\documentclass[12pt]{article}
\usepackage{a4wide}
\usepackage{epsfig}
\usepackage{amsmath}

\newlength{\absize}
\catcode`@=11
\def\citer{\@ifnextchar [{\@tempswatrue\@citexr}{\@tempswafalse\@citexr[]}}

\def\@citexr[#1]#2{\if@filesw\immediate
  \write\@auxout{\string\citation{#2}}\fi
  \def\@citea{}\@cite{\@for\@citeb:=#2\do
    {\@citea\def\@citea{--\penalty\@m}\@ifundefined
       {b@\@citeb}{{\bf ?}\@warning
       {Citation `\@citeb' on page \thepage \space undefined}}%
\hbox{\csname b@\@citeb\endcsname}}}{#1}} \catcode`@=12
\begin{document}
\begin{center}
{\large\bf Commutator Anomaly in Noncommutative  Quantum Mechanics}
\vskip 1cm Sayipjamal Dulat$^{a,c,}$\footnote{sdulat@xju.edu.cn} and
Kang Li$^{b,c,}$\footnote{kangli@hztc.edu.cn}\\\vskip 1cm

{\it\small$^a$Department of Physics, Xinjiang University, Urumqi
830046
China\\
$^b $Department of Physics, Hangzhou Teachers College,Hangzhou,
310036, China\\
$^c $ The Abdus Salam International Centre for Theoretical Physics,
Trieste, Italy}


\vskip 0.5cm
\end{center}

\begin{abstract}
In this letter, firstly, the Schr$\ddot{o}$dinger equation on
noncommutative phase space is given by using a generalized Bopp's
shift. Then the anomaly term of commutator of arbitrary physical
observable operators on noncommutative phase space is obtained.
Finally, the  basic uncertainty relations for  space-space and
space-momentum  as well as  momentum-momentum operators in
noncommutative quantum mechanics (NCQM), and uncertainty relation
for arbitrary physical observable operators in NCQM are discussed.
\end{abstract}

\noindent\vskip 0.2cm {\small{\bf Keywords:} Noncommutative phase
space,
Commutator anomaly, Uncertainty relations. }

\vskip 0.5cm
 Recently, there has been much interest in the
study of physics on noncommutative(NC) space\cite{SW}-\cite{Scho},
not only because the NC space is necessary when one studies the low
energy effective theory of D-brane with B field background, but also
because in the very tiny string scale or at very high energy
situation, the effects of non commutativity of both space-space and
momentum-momentum may appear. There are many papers devoted to the
study of various aspects of noncommutative quantum mechanics
\cite{DJT}-\cite{Likang}. For example, the Schr$\ddot{o}$dinger
equation on noncommutative space is discussed in many papers
\cite{GLMR2,CST,kochan} and the uncertainty relations on
noncommutative space are discussed in\cite{CZ,BK}. However, the
study of the quantum theory on noncommutative space is far from
complete, there are a lot of physical concepts need to make clear.

In this letter,  first of all, we give the Schr$\ddot{o}$dinger
equation on noncommutative phase space by using the generalized
Bopp's shift. Besides, we calculate the anomaly term of commutator
of arbitrary physical observable operators on noncommutative phase
space. We also present a few examples to illustrate this anomaly.
Finally , we discuss the uncertainty relations of physical
observable operators in NCQM.

In the usual $n$ dimensional commutative space , the coordinates and
momenta in quantum mechanics have the following commutation
relations:

\begin{eqnarray}
\label{Eq:cmr}
 \begin{array}{l}
~[x_{i},x_{j}]=0,\\~ [p_{i},p_{j}]=0,\hspace{2cm} i,j = 1,2,
...,n\\~ [x_{i},p_{j}]=i \hbar\delta_{ij}.
\end{array}
\end{eqnarray}
At very tiny scales, say string scale,  not only space-momentum
does not commute, but also both space-space and momentum-momentum
may not commute anymore. Let us denote the operators of
coordinates and momenta on noncommutative phase space as $\hat{x}$
and $\hat{p}$ respectively, then  $\hat{x}_i$ and $\hat{p}_i$ will
have the following algebraic relations\cite{Likang}, if both
space-space and momentum-momentum non-commutativity are considered
\begin{equation}
\label{Eq:nmr}
 \begin{array}{l}
~[\hat{x}_{i},\hat{x}_{j}]=i\Theta_{ij}, \\~
[\hat{p}_{i},\hat{p}_{j}]=i\bar{\Theta}_{ij},\hspace{2cm} i,j =
1,2,...,n\\~ [\hat{x}_{i},\hat{p}_{j}]=i \hbar\delta_{ij}.
\end{array}
\end{equation}
Here $\{\Theta_{ij}\}$ and $\{\bar{\Theta}_{ij}\}$  are totally
antisymmetric matrices with very small elements representing the
noncommutative property of the space and momentum on
noncommutative phase space. The representations  of the
noncommutative $\hat{x}$ and $\hat{p}$ in term of $x$ and $p$ were
given in ref.\cite{Likang}. So that the noncommutative problems
can be changed into problems in the usual commutative space which
we are familiar with
\begin{equation}
\label{Eq:Rep.4}
 \begin{array}{ll}
 \hat{x}_{i}&= \alpha x_{i}-\frac{1}{2\hbar\alpha}\Theta_{ij}p_{j},\\
 ~&~\\
 \hat{p}_{i}&=\alpha p_{i}+\frac{1}{2\hbar\alpha}\bar{\Theta}_{ij}x_{j},\hspace{1cm} i,j =
 1,2,...,n.
\end{array}
\end{equation}
Here $\alpha$ is a scaling constant related to the
noncommutativity of phase space.When $\bar{\Theta}=0$, it leads
$\alpha =1$\cite{Likang}, which is  the case that is extensively
studied in the literature, where the space-space is non-commuting
while momentum-momentum is commuting.

The Schr$\ddot{o}$dinger equation on noncommutative phase space is
usually expressed as
\begin{equation}
i\hbar\frac{\partial}{\partial t}\psi =H(x,p)\ast\psi ,
\end{equation}
where $H(x,p)$ is the Hamiltonian operator in quantum mechanics,
and the $\ast$ product is the well know Moyal-Weyl product. On
noncommutative phase space the $\ast$ product can be replaced by a
generalized Bopp's shift
\begin{equation}\label{GBShift}
\begin{array}{ll}
i\hbar\frac{\partial}{\partial t}\psi &=H(\hat{x}_i,\hat{p}_i)\ast\psi\\
~&=H(\alpha x_{i}-\frac{1}{2\hbar\alpha}\Theta_{ij}p_{j},\alpha
p_{i}+\frac{1}{2\hbar\alpha}\bar{\Theta}_{ij}x_{j})\psi .
\end{array}
\end{equation}
When $n=2, \bar{\Theta}_{ij}=0$, our results reduce to the the
Bopp's shift\cite{CFZ} $i\hbar\frac{\partial}{\partial t}\psi =H(
x_i-\frac{1}{2}\Theta \epsilon_{ij} p_j) \psi$  which is discussed
in many papers. From equation (\ref{GBShift}), we can see that the
Hamiltonian operator on NC phase space can be defined as
\begin{equation}
\hat{H}=H(\hat{x}_i,\hat{p}_i)=H(\alpha
x_{i}-\frac{1}{2\hbar\alpha}\Theta_{ij}p_{j},\alpha
p_{i}+\frac{1}{2\hbar\alpha}\bar{\Theta}_{ij}x_{j}).
\end{equation}
This give us a hint of how to define an arbitrary physical
observable operator on NC phase space. If a physical observable
operator in usual quantum mechanics has the form
\begin{equation}
A=A(x_i, p_i),
\end{equation}
then the same physical observable operator on NC phase space is
defined as
\begin{equation}
\hat{A}=A(\hat{x}_i, \hat{p}_i).
\end{equation}
In quantum mechanics, if the commutator of two observable operators
$A$ and $B$ is $[A,B]=iC$, then in NCQM, the commutator
$[\hat{A},\hat{B}]=i\hat{C}$ does not hold anymore because of the
existence of the anomaly terms. By using Taylor's expansion, the
physical operator $\hat A$ on NC phase space can be expanded as
\begin{eqnarray}
\hat A \equiv A(\hat{x}_i,\hat{p}_i)&=& A(\alpha x_i,\alpha p_{i})
-\frac{1}{2\hbar\alpha}\Theta_{nm}p_{m}\frac{\partial
{\hat{A}}}{\partial \hat{x}_n}|_{\Theta =\bar \Theta=0}\nonumber\\
&+&\frac{1}{2\hbar\alpha}\bar{\Theta}_{nm}x_{m}\frac{\partial
{\hat{A}}}{\partial \hat{p}_n}|_{\Theta =\bar \Theta=0} + .....
\end{eqnarray}
Then, by a tedious but straightforward  calculation, we obtain the
following commutator of  two arbitrary physical observable operators
$\hat A$ and $\hat B$ on NC phase space
\begin{eqnarray}\label{AB}
[\hat{A},\hat{B}]=i\hat{C} &+& \frac{i}{2} \Theta_{nm}\Big(
\frac{\partial {\hat{A}}}{\partial \hat{x}_n} \frac{\partial
{\hat{B}}}{\partial \hat{x}_m} +\frac{\partial {\hat{B}}}{\partial
\hat{x}_m} \frac{\partial {\hat{A}}}{\partial \hat{x}_n}
\Big)|_{\Theta =\bar \Theta=0} \nonumber\\&+& \frac{i}{2}\bar
\Theta_{nm}\Big( \frac{\partial {\hat{A}}}{\partial \hat{p}_n}
\frac{\partial {\hat{B}}}{\partial \hat{p}_m} +\frac{\partial
{\hat{B}}}{\partial \hat{p}_m} \frac{\partial {\hat{A}}}{\partial
\hat{p}_n} \Big)|_{\Theta =\bar \Theta=0}+O(\Theta^2),
\end{eqnarray}
 where  $O(\Theta^2)$ stands for the second and higher order terms in $\Theta$ and $\bar
 \Theta$;
the second and third terms of r.h.s. of  (\ref{AB})  are purely
noncommutative effects of the first order on the phase space, and we
call them commutator anomaly on NC phase space.

For example, by (\ref{AB}), one can immediately get the following
commutation relations between $\hat{\mathbf{L}}$ and
$\hat{\mathbf{x}},\hat{\mathbf{p}}$
\begin{equation}\label{Eq:cr1}
 [\hat{L}_{i},\hat{x}_{j}]= i\hbar\epsilon_{ijk}\hat{x}_{k}+i\epsilon_{ikl}\Theta_{kj}\hat{p}_{l},
\end{equation}
\begin{equation}\label{Eq:cr2}
 [\hat{L}_{i},\hat{p}_{j}]=i\hbar\epsilon_{ijk}\hat{p}_{k}-i\epsilon_{ikl}\bar{\Theta}_{kj}\hat{x}_{l}.
\end{equation}
Which coincide with the direct calculation in ref.\cite{Likang},
and where the angular momentum on noncommutative  phase space has
the following form
\begin{equation}\label{orbital}
\hat{L}_{i}=\epsilon_{ijk}\hat{x}_{j}\hat{p}_{k}.
\end{equation}

As an another example of (\ref{AB}), let's consider  the harmonic
oscillator on $n$ dimensional noncommutative phase space, the
Hamiltonian  has the form as
\begin{equation}
\label{H1}
\hat{H}=\frac{1}{2\mu}\hat{p_{i}}\hat{p}_{i}+\frac{1}{2}\mu\omega^2\hat{x}_{i}\hat{x}_{i},
\end{equation}
where the $\mu$ and $\omega$ represent the mass and angular
frequency of the oscillator.  In quantum mechanics, we know that
the Hamiltonian operator $H$ and the angular momentum operator
$L_i$ are commute. But in NCQM, they do not commute anymore. By
using (\ref{AB}), the commutator of the two operators $\hat{H}$ in
(\ref{H1}) and $\hat{L}_i$ in (\ref{orbital}) is written as
\begin{equation}\label{HL}
 [\hat{H},\hat{L}_{i}] =
 \frac{i}{2}\mu\omega^2\Theta_{lj}\epsilon_{ijk}(\hat{x}_l\hat{p}_k
 +\hat{p}_k\hat{x}_l)+\frac{i
}{2\mu} \bar{\Theta}_{lk}\epsilon_{ijk}(\hat{x}_j\hat{p}_l
 +\hat{p}_l\hat{x}_j).
\end{equation}

Usual uncertainty principle in quantum mechanics, the so called
Heisenberg uncertainty principle, should be reformulated because of
the noncommuting nature of both space-space and momentum-momentum
operators in NCQM. Therefore, as a final step, we discuss the
uncertainty relations implied by the basic commutation relations
(\ref{Eq:nmr}) and generalized commutator (\ref{AB}) of two
observable operators in NCQM.

In usual quantum mechanics the uncertainty relation for noncommuting
arbitrary physical quantities corresponding to the operators $A$ and
$B$ is
\begin{equation}
\Delta A_\psi\cdot\Delta B_\psi\geq\frac{1}{2}|\langle
C\rangle_\psi|,
\end{equation}
for any normalized state $\psi$; here
\begin{equation}
\Delta A_\psi\equiv\sqrt{\displaystyle(\psi,(A-\langle A\rangle_\psi
I)^2\psi)}, \text{\ etc,} \label{w9}
\end{equation}
where $\bf I$ is the identity operator. From equation
(\ref{Eq:nmr}), it is easy to get the following basic uncertainty
relations on NC phase space:
\begin{subequations}
\label{w12}
\begin{align}
\Delta \hat{x}_i\Delta \hat{x}_j\geq\frac{1}{2} \Theta_{ij},\label{w12a}\\
\Delta \hat{x}_i\Delta \hat{p}_j\geq\frac{1}{2}\hbar \delta_{ij},\label{w12b}\\
\Delta \hat{p}_i\Delta
\hat{p}_j\geq\frac{1}{2}\bar\Theta_{ij},\label{w12c}
\end{align}
\end{subequations}
for   $i,j=1,2\ldots n$. From Eq. (\ref{AB}), the uncertainty
relation of two arbitrary observable operators in the first order of
$\Theta$ and $\bar{\Theta}$, on NC phase space, for any normalized
state $\psi$, has the form
\begin{eqnarray}\label{w8}
\Delta \hat{A}_\psi\cdot\Delta
\hat{B}_\psi&\geq&\frac{1}{2}|\langle\hat{C}\rangle_\psi| +
{\frac{1}{2}}\Big|\Big\langle \Big( \Theta_{nm}\frac{\partial
{\hat{A}}}{\partial \hat{x}_n} \frac{\partial {\hat{B}}}{\partial
\hat{x}_m}+ \bar{\Theta}_{nm} \frac{\partial {\hat{A}}}{\partial
\hat{p}_n} \frac{\partial {\hat{B}}}{\partial \hat{p}_m}
 \Big)|_{\Theta =\bar \Theta=0}
\Big\rangle_\psi \Big|,
\end{eqnarray}
 obviously it holds for all physical quantities with noncommuting operators. Here
\begin{equation}
\Delta\hat{ A}_\psi\equiv\sqrt{\displaystyle(\psi,(\hat{A}-\langle
\hat{A}\rangle_\psi I)^2\psi)}, \text{\ etc.} \label{w9}
\end{equation}
For example, for the noncommutative harmonic oscillator system, by
 (\ref{HL}) we have the following uncertainty relation for
$\hat{H}$ and $\hat{L}_i$
\begin{equation}
\triangle\hat{H}\cdot\triangle\hat{L_i}\geq
\frac{1}{2}\mu\omega^2\Theta_{lj}\epsilon_{ijk}\langle\hat{x}_l\hat{p}_k\rangle_\psi
 +\frac{1 }{2\mu}
 \bar{\Theta}_{lk}\epsilon_{ijk}\langle\hat{x}_j\hat{p}_l\rangle_\psi.
\end{equation}

When $\bar{\Theta}_{ij}=0$, which led to $\alpha=1$, in this case
the momentum-momentum are no more noncommutative. Thus from
$(\ref{w12})$ one writes the following basic non-vanishing
uncertainty relations
\begin{subequations}
\label{w13}
\begin{align}
\Delta \hat{x}_i\Delta \hat{x}_j\geq\frac{1}{2} \Theta_{ij},\label{w13a}\\
\Delta \hat{x}_i\Delta \hat{p}_j\geq\frac{1}{2}\hbar
\delta_{ij},\label{w13b}
\end{align}
\end{subequations}
which have been discussed in Ref.\cite{BK}. Also by $(\ref{w8})$,
the uncertainty relation of two arbitrary observable operators in
the first order of $\Theta$ has the form
\begin{eqnarray}\label{w14}
\Delta \hat{A}_\psi\cdot\Delta
\hat{B}_\psi&\geq&\frac{1}{2}|\langle\hat{C}\rangle_\psi| +
{\frac{1}{2}}\Theta_{nm}\Big|\Big\langle  \frac{\partial
{\hat{A}}}{\partial \hat{x}_n} \frac{\partial {\hat{B}}}{\partial
\hat{x}_m} | _{\Theta =0} \Big\rangle_\Psi \Big|.
\end{eqnarray}
We can see from (\ref{w14}) that even for space-space noncommuting
and momentum-momentum commuting case $(\alpha
=1,\bar{\Theta}_{ij}=0)$ ,
 the commutator anomaly remains there.

In summary,  At first, we give out the Schr$\ddot{o}$dinger equation
on NC phase space by introducing a generalized Bopp's shift. Then by
defining an arbitrary physical observable on NC phase space, we
calculate  the general commutator anomaly for arbitrary physical
observables. At last, we give the uncertainty relations on NC phase
space. A few  remarks on the basic uncertainty relation (\ref{w12})
is in order: in noncommutative phase space, the three basic types of
uncertainty relations (\ref{w12}) can not reach the minimal
uncertainty simultaneously. It may related to three kinds of
coherent states which will be discussed in our forthcoming paper.

\subsection*{Acknowledgements}
This letter is completed during our visit to the high energy
section of the Abdus Salam International Centre for Theoretical
Physics (ICTP). We would like to thank Prof. S. Randjbar-Daemi for
his kind invitation and warm hospitality during our visit at the
ICTP. This work is supported in part by the National Natural
Science Foundation of China (90303003 and 10465004) and the
Natural Science Foundation of Zhejiang Provence, China
(M103042,102011,102028). The authors also recognize the support
from the  Abdus Salam ICTP, Trieste, Italy.

\end{document}